\documentclass{article}
\usepackage{ijcai19}

\usepackage{times}
\usepackage{soul}
\usepackage{url}
\usepackage[hidelinks]{hyperref}
\usepackage[utf8]{inputenc}
\usepackage[small]{caption}
\usepackage{graphicx}
\usepackage{amsmath}
\usepackage{amssymb}
\usepackage{booktabs}
\usepackage{algorithm}
\usepackage{algpseudocode}

\usepackage[T1]{fontenc}
\title{Single-Channel Signal Separation and Deconvolution \\ with Generative Adversarial Networks}

\author{
Qiuqiang Kong$^1$
\and
Yong Xu$^2$
\and
Wenwu Wang$^1$
\and
Philip J.B. Jackson$^1$
\And
Mark D. Plumbley$^1$
\affiliations
$^1$University of Surrey, Guildford, UK\\
$^2$Tencent AI lab, Bellevue, USA
\emails
\{q.kong, w.wang, p.jackson, m.plumbley\}@surrey.ac.uk, lucayongxu@tencent.com
}

\begin{document}

\maketitle

\begin{abstract}
Single-channel signal separation and deconvolution aims to separate and deconvolve individual sources from a single-channel mixture and is a challenging problem in which no prior knowledge of the mixing filters is available. Both individual sources and mixing filters need to be estimated. In addition, a mixture may contain non-stationary noise which is unseen in the training set. We propose a synthesizing-decomposition (S-D) approach to solve the single-channel separation and deconvolution problem. In synthesizing, a generative model for sources is built using a generative adversarial network (GAN). In decomposition, both mixing filters and sources are optimized to minimize the reconstruction error of the mixture. The proposed S-D approach achieves a peak-to-noise-ratio (PSNR) of 18.9 dB and 15.4 dB in image inpainting and completion, outperforming a baseline convolutional neural network PSNR of 15.3 dB and 12.2 dB, respectively and achieves a PSNR of 13.2 dB in source separation together with deconvolution, outperforming a convolutive non-negative matrix factorization (NMF) baseline of 10.1 dB. 
\end{abstract}

\section{Introduction}

Single-Channel signal separation and deconvolution aims to separate and deconvolve sources from a single-channel mixture. One challenging aspect of single-channel signal separation and deconvolution is that only a single-channel mixture is available, so this problem is underdetermined. Second, there is no prior knowledge of the mixing filters. Both individual sources and mixing filters are unknown and need to be estimated. Third, there is no prior knowledge on the noise, which can be non-stationary and has not been seen in the training data. These difficulties lead to single-channel signal separation and deconvolution being a very challenging problem. Single-channel signal separation and deconvolution has many applications in image, speech and audio denoising \cite{xie2012image}, inpainting \cite{yeh2016semantic}, deconvolution and separation 
\cite{cichocki2009nonnegative,mijovic2010source}. For example, an audio sensor usually receives signals from multiple sources convolved with channel distortion. 

Much previous work focuses on source separation \cite{cichocki2009nonnegative,grais2014deep} or deconvolution \cite{levin2009understanding,campisi2017blind} independently, but not together. We categorize previous source separation and deconvolution methods into decomposition based approaches and regression based approaches. Decomposition methods usually learn a set of bases for sources and use these bases to decompose a mixture. Decomposing methods including non-negative matrix factorization (NMF) \cite{lee1999learning,cichocki2009nonnegative,kitamura2013music} assumes that %that assumes ~ yong xu
a source can be represented by linear combination of a set of bases. NMF has been used in source representation and separation \cite{cichocki2006new,kitamura2013music}. In contrast to the decomposition based approaches, regression based approaches learn a mapping from a mixture to an individual source. Such mappings can be modeled by neural networks, for example, fully connected neural networks \cite{grais2014deep} and convolutional neural networks (CNNs) \cite{jain2009natural,zhang2017beyond}. In \cite{xie2012image}, a stacked denoising auto-encoder (DAE) is proposed to recover sources from a mixture. CNNs are used for source deconvolution in \cite{xu2014deep}.

However, many decomposition methods such as NMF and ICA are shallow layer models, which are typically a linear combination of bases. These shallow layer models do not have enough capacity to represent a broad range of sources compared with neural networks \cite{jain2009natural}. On the other hand, regression based approaches such as deep neural networks are able to model complicated mappings but require both mixture and target sources for training. Regression based methods may not generalize well if the mixing filter and noise in the testing data have different distribution from the training data, which will result in poor separation results when the mixing filter and noise are unseen in the training data \cite{yosinski2014transferable}. Recently generative adversarial networks (GANs) have been proposed for solving the source separation problem \cite{fan2017svsgan,subakan2017generative,stoller2017adversarial}. So far these methods assume % assumes ~ yongxu
that the mixing filters in the single-channel signal separation problem are known. 

This paper proposed a novel synthesizing-decomposition (S-D) approach to solve the single-channel source separation and deconvolution problem. Compared to the conventional regression approaches, the S-D approach applies generative adversarial network (GANs) to solve this problem in a generative way. The S-D approach can estimate both the sources and convolutive mixing filters, while conventional regression methods do not estimate convolutive mixing filters. In addition, we formulate the single-channel signal separation and deconvolution problem as a Bayesian maximum a posteriori (MAP) estimation which is a constrained non-convex optimization problem. In the S-D approach, a generative model is built for sources using a generative adversarial network (GAN). In decomposition, both sources and mixing filters can be obtained by minimizing the reconstruction error of a mixture. To tackle the non-convex optimization problem, repeating the decomposition with different initializations can significantly increase the underdetermined single-channel signal separation and deconvolution performance. We carry out the underdetermined single-channel signal separation and deconvolution experiments on MNIST dataset as a starting research to show the effectiveness of the proposed S-D approach with GANs. 

This paper is organized as follows: Section 2 formulates the underdetermined single-channel signal separation and deconvolution problem. Section 3 proposes the synthesising-decomposition (S-D) approach for this problem. Section 4 shows experimental results. Section 5 concludes and forecasts future work. 

\section{Single-Channel Signal Separation and Deconvolution}

In underdetermined single-channel signal separation and deconvolution, a single-channel mixture $ x(u) \in L^{2}(\Omega), u \in \Omega $ is composed of individual sources $ s_{k}(u) \in L^{2}(\Omega), u \in \Omega, k=1, ..., K $ convolved with unknown filters $ \alpha_{k}(u)  \in L^{2}(\Omega), u \in \Omega, k=1, ..., K $ followed by unknown additional noise $ n(u) \in L^{2}(\Omega), u \in \Omega $. The space $ \Omega $ can be a Euclidean space $ \mathbb{R}^{d} $ where $ K $ and $ d $ denote the number and the dimension of sources, respectively:
\begin{equation} \label{eq:defination}
x(u) = \sum_{k=1}^{K} (\alpha_{k} \ast s_{k})(u) + n(u). 
\end{equation}
The symbol $ \ast $ represents the convolution operation:
\begin{equation} \label{eq:convolution}
(\alpha_{k} \ast s_{k})(u) = \int_{\mathbb{R}^{d}} \alpha_{k}(u - v)s_{k}(v)dv .
\end{equation}
For the simple case of source separation without deconvolution, in (\ref{eq:convolution}) $ \alpha_{k}(u) $ simplifies to $ \alpha_{k}(u-v) = \alpha_{k} \delta(u-v) $ where $ \delta(u) $ is the Dirac delta function. General single-channel signal separation and deconvolution problem concerns both separating and deconvolving individual sources $ s_{k}(u), k=1, ..., K $ from a single-channel mixture $ x(u) $ while the mixing filters $ \alpha_{k}(u), k=1, ..., K $ and the noise signal $ n(u) $ are unknown in (\ref{eq:defination}). In the following paper, we simplify the notation of $ x(u), s_{k}(u), \alpha_{k}(u) $ to $ x, s_{k} $ and $ \alpha_{k} $, respectively. 

In the regression based approaches \cite{jain2009natural,grais2014deep}, a mapping from a mixture to a source signal is modeled by deep neural networks and learned to separate the $ k $-th source: $ f_{k}: x \mapsto s_{k} $. In separation, separated sources are obtained by forwarding a mixture to the model: $ \hat{s}_{\text{test}} = f_{k}(x_{\text{test}}) $. However there are several problems associated with the regression based approaches as follows:

\paragraph{Problem 1.} In regression based supervised learning, the training data $ x_{\text{train}} $ and testing data $ x_{\text{test}} $ should have the same distribution, otherwise the trained model will be biased \cite{yosinski2014transferable}. However, in single-channel signal separation and deconvolution, no prior knowledge of test noise $ n $ is available. The model trained with training noise may not generalize well to sources with unseen non-stationary noise.

\paragraph{Problem 2.} In single-channel signal separation and deconvolution, both the sources $ s_{k} $ and mixing filters $ \alpha_{k} $ are unknown and need to be estimated.

\paragraph{Problem 3.} Previous regression and decomposition based approaches do not constrain the distribution of the separated sources $ \hat{s} $ to be the same as the distribution of real sources $ p_{\text{real}}(s) $. Ideally, the separated sources $ \hat{s} $ should be regularized in the area where $ p_{\text{real}}(\hat{s}) $ has larger value. 

Decomposition approaches such as NMF can be trained on individual sources instead of on a mixture so that Problem 1 can be mitigated. Recently, GANs \cite{fan2017svsgan,subakan2017generative,stoller2017adversarial} have been applied to source separation to solve Problem 3 to constrain the separated sources to be laid in natural source space. However, those methods are based on the assumption that the mixing filters $\alpha_{k}$ are constants so that they are solving only separation but not deconvolution problem as shown in (1). 

\section{Proposed Synthesising-Decomposition (S-D) Approach}

\subsection{Maximum a Posteriori (MAP) Estimation}

In this section, we first formulate the single-channel signal separation and deconvolution problem in (\ref{eq:defination}) as a Bayesian parameter estimation problem. We denote $ \theta = \{ s_{1}, ..., s_{K}, \alpha_{1}, ..., \alpha_{K} \} $ as the set of parameters to be estimated, including sources and mixing filters. The estimated $ \hat{\theta} $ can be obtained by maximum a posteriori (MAP) estimation:
\begin{equation} \label{eq:argmax}
\begin{split}
\hat{\theta} &= \underset{\theta}{\mathrm{argmax}} \ p(\theta | x) \\
& = \underset{\theta}{\mathrm{argmax}} \ p(x | \theta) p(\theta). 
\end{split}
\end{equation}
The first term $ p(x|\theta) $ in (\ref{eq:argmax}) is a likelihood function. The reconstructed signal can be written as $ \hat{x} = \sum_{k=1}^{K}\alpha_{k} \ast s_{k} $. Assuming $ n $ is a Gaussian process, the likelihood of observed signal given estimated signal can be written as:
\begin{equation} \label{eq:prob_s_given_theta}
\begin{split}
p(x|\theta) & = p(x|\hat{x}) \\
& = \prod_{u \in \Omega} p(x(u)|\hat{x}(u)) \\
& = \prod_{u \in \Omega} \mathcal{N}(x(u) - \sum_{k=1}^{K} \alpha_{k} \ast s_{k}, \sigma_{n})
\end{split}
\end{equation}
\noindent where $ \mathcal{N}(\cdot, \cdot) $ is the probability density of a Gaussian distribution. The second term $ p(\theta) $ in (\ref{eq:argmax}) is the prior probability of $ \theta $. Assuming the sources and filters are independent of each other, we can write $ p(\theta) $ as:
\begin{equation} \label{eq:prob_theta}
p(\theta) = \prod_{k=1}^{K} p(\alpha_{k}) \prod_{k=1}^{K} p(s_{k}). 
\end{equation}
We assume $ s_{k}, k=1,...,K $ to have a compact support $ V \in \Omega $. Substituting equations (\ref{eq:prob_s_given_theta}) and (\ref{eq:prob_theta}) to equation (\ref{eq:argmax}) the estimation of sources and filters can be obtained by solving the following optimization problem:
\begin{equation} \label{eq:opt_on_x_alpha}
\begin{split}
&\hat{s}_{k}, ..., \hat{s}_{K}, \hat{\alpha}_{1}, ..., \hat{\alpha}_{K} = \underset{\substack{s_{1} \in V, ..., s_{K} \in V \\[0ex] \alpha_{1}, ..., \alpha_{J}}}{\mathrm{argmax}} \\ & \prod_{u \in \Omega} \mathcal{N}(x(u) - \sum_{k=1}^{K} \alpha_{k} \ast s_{k}, \sigma_{n}) \prod_{k=1}^{K} p(\alpha_{k}) \prod_{k=1}^{K} p(s_{k})
\end{split}
\end{equation}
\begin{algorithm}[t]
	\caption{Training of a GAN \protect\cite{goodfellow2014generative}.}\label{alg:synthesis}
	\begin{algorithmic}[1]
		\State Inputs: Real data $ s_{n}, n=1, ..., N $. 
	    \State Outputs: Parameters of the discriminator $ \theta_{d} $ and the generator $ \theta_{g} $ of a GAN. 
	    \For{number of iterations}
	        \begin{itemize}
	            \item Sample minibatch of m noise samples $ \{z^{(1)}, ..., z^{(m)}\} $ from a Gaussian distribution $ N(0, 1) $. 
                \item Sample minibatch of m examples $ \{s^{(1)}, ..., s^{(m)}\} $ from real data. 
                \item Update the discriminator by ascending its stochastic gradient: 
                \begin{equation*}
                \triangledown \theta_{d} \frac{1}{m} \sum_{i=1}^{m} \left [ \text{log}D(s^{(i)}) + \text{log}(1 - D(G(z^{(i)}))) \right ]
                \end{equation*}
                \item Sample minibatch of m noise samples $ \{z^{(1)}, ..., z^{(m)}\} $ from a Gaussian distribution $ N(0, 1) $.
                \item Update the generator by descending its stochastic gradient:
                \begin{equation*}
                \triangledown \theta_{g} \frac{1}{m} \sum_{i=1}^{m} \text{log}(1 - D(G(z^{(i)})))
                \end{equation*}
                 \end{itemize}
		\EndFor
	\end{algorithmic}
\end{algorithm}

\begin{algorithm}[t]
	\caption{Decomposition of a mixture source. Hyperparameters: $K$: Number of individual sources. }\label{alg:decomposition}
	\begin{algorithmic}[1]
		\State Inputs: A mixture source. Generator $ G $ trained using algorithm \ref{alg:synthesis}. 
	    \State Outputs: Separated and deconvolved sources $ s_{k}, k=1, ..., K $ and mixing filters $ \alpha_{k}, k=1, ..., K $. 
	    \State Sample $ K $ seeds $ \{z_{1}, ..., z_{K} \} $ and $ K $ mixing filters $ \{\alpha_{1}, ..., \alpha_{K}\} $ from a Gaussian distribution $ N(0, 1) $. 
	    \For{number of iterations}
	        \begin{itemize}
	            \item Calculate reconstructed signal $ \hat{s} = \sum_{k=1}^{K} \alpha_{k} \ast G(z_{k}) $. 
	            \item Calculate gradient $ \triangledown_\phi $ from equation (\ref{eq:gradient}) where $ \phi = \{z_{1}, ..., z_{K}, \alpha_{1}, ..., \alpha_{K}\} $.
	            \item Update $ \phi = \{z_{1}, ..., z_{K}, \alpha_{1}, ..., \alpha_{K}\} $ using Adam optimizer \cite{kingma2014adam}. 
                 \end{itemize}
		\EndFor
	\end{algorithmic}
	
\end{algorithm}

\subsection{Optimization with S-D Approach}

To optimize (\ref{eq:opt_on_x_alpha}) is difficult because of the constraint of $ s_{k} \in V $. The source prior $ p(s_{k}) $ is unknown, so that $ V $ can not be written in a closed form. Our solution is to convert (\ref{eq:opt_on_x_alpha}) to an unconstrained optimization problem. In the proposed S-D approach, we first build a generative model for $ x_{k} $ with a GAN \cite{goodfellow2014generative,subakan2017generative}. A GAN consists of a generator $ G $ and a discriminator $ D $. The generator $ G $ is a mapping from any distribution $ p_{z} $ such as a Gaussian distribution $ N(0, \sigma \mathbf{I}) $ to a real distribution of sources. We call $ p_{z} $ a \textit{seed distribution} and sample $ z \sim p_{z} $ as seeds. The generator $ G $ is trained to generate samples to fool the discriminator $ D $. The discriminator $ D $ is trained to discriminate fake sources from real sources. In other words, the generator $ G $ and the discriminator $ D $ play the following two-player minimax game with value function $ V(G, D) $ \cite{goodfellow2014generative}:
\begin{equation} \label{error_rate}
\begin{split}
\underset{G}{\text{min}} \ \underset{D}{\text{max}} V(D, G) = \ \ \ \ \ \ \ \ \ \ \ \ \ \ \ \ \ \ \ \ \  \ \ \ \ \ \ \ \ \ \ \ \ \ \ \ \ \ \ \ \ \ \ \ \ \ \ \ \ \ \ \ \\ \mathbb{E}_{s \sim p_{\text{data}}(s)}[\text{log} D(x)] + \mathbb{E}_{z \sim p_{z}(z)}[ \text{log}(1 - D(G(z))) ]
\end{split}
\end{equation}
where $ p_{\text{data}} $ is the real data probability density. The training of the GAN is shown in Algorithm 1. The generator $ G $ and discriminator $ D $ are trained iteratively. If both $ G $ and $ D $ have enough capacity, then the generated source distribution will converge to $ p_{\text{data}} $ \cite{subakan2017generative}. Once GAN is successfully trained, there is $ G(z) \in V $ for all $z$. To solve the optimization problem in (\ref{eq:opt_on_x_alpha}), we substitute $ s_{k} = G(z_{k}) $ and optimize over $ z_{k} $ instead of $ s_{k} $ so that the constraint $ s \in V $ is eliminated. Now the variables to be optimized are $ z_{k} $ and the mixing filters $ \alpha_{k} $. In addition, GAN does not predict the probability density $ p(s_{k}) $ of $ s_{k} $ so the optimization of equation (\ref{eq:opt_on_x_alpha}) is intractable. To solve this problem, we approximate $ p(s_{k}) $ with:
\begin{equation} \label{eq:probability_density}
\begin{split}
p(s_{k}) = \begin{cases} 0, & s_{k} \notin V \\ 1 / \left | V \right |, & s_{k} \in V. \end{cases}
\end{split}
\end{equation}
Equation (\ref{eq:probability_density}) assumes the probability density $ p(s_{k}) $ outside $ V $ is zero. It is not required to know the value of $ \left | V \right | $ as it is eliminated when optimizing (\ref{eq:opt_on_x_alpha}):

\begin{equation} \label{eq:opt_on_z_alpha}
\begin{split}
& \hat{z}_{k}, ..., \hat{z}_{K}, \hat{\alpha}_{1}, ..., \hat{\alpha}_{K} = \underset{\substack{{z}_{1}, ..., {z}_{K}\\[0ex] \alpha_{1}, ..., \alpha_{J}}}{\mathrm{argmax}} \\ & \prod_{u \in \Omega} \mathcal{N}(x(u) - \sum_{k=1}^{K} \alpha_{k} \ast s_{k}, \sigma_{n}) \prod_{k=1}^{K} p(\alpha_{k}). 
\end{split}
\end{equation}

 We assume the coefficients in $ \alpha_{k} $ to be Gaussian $ \alpha_{k} \sim \mathcal{N}(0, \sigma_{\alpha}) $. Taking the logarithm of (\ref{eq:opt_on_z_alpha}) the optimization can be written as:

\begin{equation} \label{eq:final}
\begin{split}
& \hat{z}_{k}, ..., \hat{z}_{K}, \hat{\alpha}_{1}, ..., \hat{\alpha}_{K} = \underset{\substack{{z}_{1}, ..., {z}_{K}\\[0ex] \alpha_{1}, ..., \alpha_{J}}}{\mathrm{argmin}} \\ & \left \| x - \sum_{k=1}^{K} \alpha_{k} \ast G(z_{k}) \right \|_{2}^{2} + \beta \sum_{k=1}^{K} \left \| \alpha_{k} \right \|_{2}^{2} \\
\end{split}
\end{equation}
\noindent where $ \beta = \sigma_{n} / \sigma_{\alpha} $ is a regularization term for (\ref{eq:final}). 

\begin{table*}[t]
\centering
\begin{tabular}{*{3}{c}}
 \toprule
 & Noise $ n $ & Mixing filters $ \alpha_{k}, k=1, ..., K $ \\
 \midrule
 Denoising & Gaussian & $ K=1 $, $ \alpha_{k} $ is a constant \\
 Inpainting, Completion & Unknown & $ K=1 $, $ \alpha_{k} $ is a constant \\
 \midrule
 Deconvolution & - & $ K=1 $, $ \alpha_{k} $ is a tensor \\
 Separation & - & $ K>1 $, $ \alpha_{k} $ are constants \\
 Separation + deconvolution & - & $ K>1 $, $ \alpha_{k} $ are tensors \\
 \bottomrule
\end{tabular}
\caption{Category of single-channel signal separation and deconvolution problem with different noise and mixing filters. }
\label{table:categorization}
\end{table*}

\subsection{Optimization}
To solve (\ref{eq:final}), we apply a gradient based iterative approach. We denote $ \phi = \{ z_{1}, ..., z_{K}, \alpha_{1}, ..., \alpha_{K} \} $ where $ z_{k} $ and $ \alpha_{k} $ need to be optimized. First we randomly initialize $ \phi $, then the gradients of $ \phi $ are calculated by:
\begin{equation} \label{eq:gradient}
\begin{split}
\triangledown_{\phi} = \frac{\partial }{\partial \phi} \left \{ \left \| x - \sum_{k=1}^{K} \alpha_{k} \ast G(z_{k}) \right \|_{2}^{2} + \beta \sum_{k=1}^{K} \left \| \alpha_{k} \right \|_{2}^{2} \right \}. 
\end{split}
\end{equation}

The parameters $ \phi $ are optimized using Algorithm 2. Because $ G $ is a non-linear mapping, so (\ref{eq:final}) is a non-convex function over $ \phi $. The gradient based methods might reach a local minimum depending on the initialization of seeds. To mitigate this problem we repeat Algorithm \ref{alg:decomposition} for $ L $ times and choose the one with smallest reconstruction error.

\section{Experiments}
In this section, we apply the proposed S-D method to solve underdetermined image single-channel signal separation and deconvolution problem. We carry out experiments on MNIST 10-digit dataset \cite{lecun1998gradient} as a starting research for this challenging problem and show the effectiveness of the proposed S-D method. With different types of unknown mixing filters $ \alpha_{k} $ and unknown interference noise $ n $, the problem of (\ref{eq:defination}) can be categorized as image denoising, inpainting, completion, deconvolution and separation, as shown in Table \ref{table:categorization}. The symbol `-' represents any type of noise. Previous works usually focus on one of these problems such as denoising \cite{jain2009natural}, inpainting \cite{xie2012image}, deconvolution \cite{xu2014deep} or separation \cite{subakan2017generative}. In this paper we solve these problem together with the proposed S-D method. The PyTorch implementation of this paper is released\footnote{https://github.com/qiuqiangkong/gan\textunderscore separation\textunderscore deconvolution}.

\subsection{Model Configuration}
In the proposed S-D approach, we model the synthesising procedure with a deep convolutive generative adversarial network (DCGAN) \cite{radford2015unsupervised}, which can stabilize the training of a GAN and can generate high quality images as shown in \cite{radford2015unsupervised}. A DCGAN consists of a generator $ G $ and a discriminator $ D $. The input to $ G $ consists of a seed sampled from a Gaussian distribution $ N(0, \sigma \mathbf{I}) $. The Gaussian distribution has a dimension of 100 following \cite{radford2015unsupervised}. The generator $ G $ has 4 transpose convolutional layers with number of feature maps of 512, 256, 128 and 1, respectively. Following \cite{radford2015unsupervised}, batch normalization \cite{ioffe2015batch} and ReLU non-linearity are applied after each transpose convolutional layer. The output of $ G $ is an image which has the same size as the images in the training data. The discriminator $ D $ takes a fake or a real image as input. The discriminator $ D $ consists of 4 convolutional layers, with a sigmoid output representing the probability that the input to $ D $ is from real data instead of generated data. Following \cite{radford2015unsupervised}, we use the Adam \cite{kingma2014adam} optimizer with a learning rate of 0.0002, a $ \beta_{1} $ of 0.5 and a $ \beta_{2} $ of 0.999 to train the generator. In decomposition, we freeze the trained generator $ G $. We approximate $ p(x|\hat{x}) $ with a Gaussian distribution which works well in our experiment. We set $ \beta $ to 0.001 to regularize the mixing filters $ \alpha_{k} $ to be searched. The filters $ \alpha_{k} $ and $ z_{k} $ are randomly initialized and optimized with Adam optimizer with a learning rate of 0.01, a $ \beta_{1} $ of 0.9 and a $ \beta_{2} $ of 0.999 (Algorithm 2).

For comparison with regression based approaches, we apply a CNN \cite{xie2012image} which consists 4 layers with batch normalization \cite{ioffe2015batch} and ReLU non-linearity. The number of layers and parameters are set to be the same as the discriminator $ D $ in the DCGAN. The CNN is trained to regress from individual source with noise $ s + n $ to individual source $ s $. For comparison with decomposition based approaches, we train a dictionary for each of the 10 digits using NMF \cite{cichocki2009nonnegative} with Euclidean distance. Each dictionary consists of 20 bases which performs well in our experiment. In decomposition, the trained dictionaries are concatenated to form a dictionary of 200 bases which is then used to decompose the mixtures.

\begin{figure*}[t] 
  \centering
  \centerline{\includegraphics[width=0.9\textwidth]{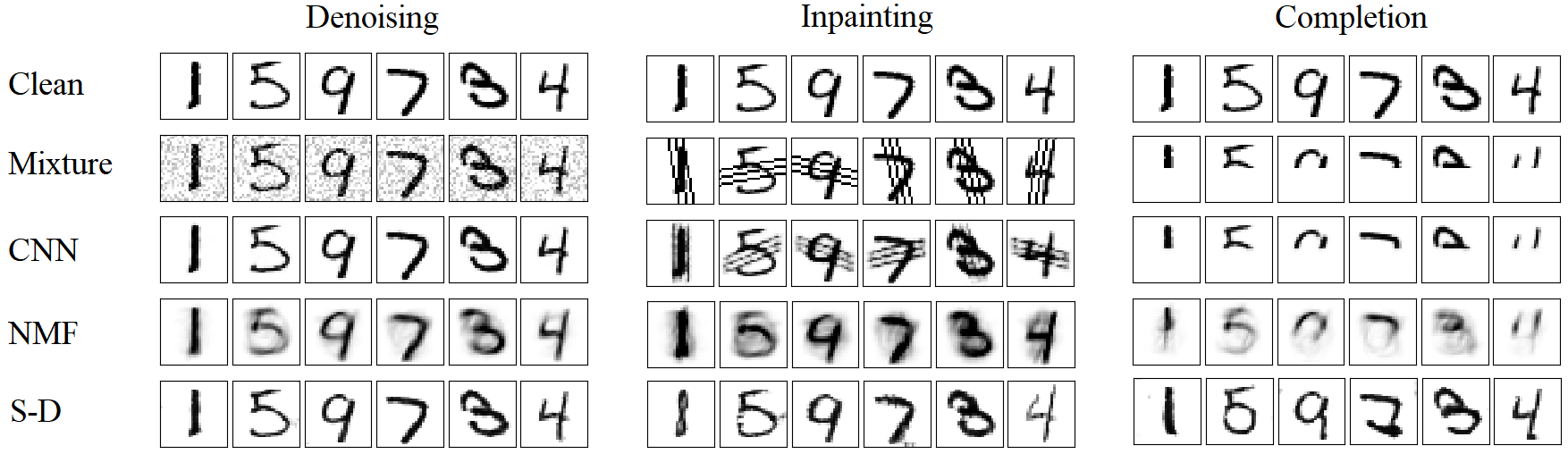}}
  \caption{Image denoising, inpainting and completion with CNN, NMF and S-D approaches. }
  \label{fig:gaussian_bar_half}
\end{figure*}

\subsection{Evaluation}
Following \cite{xie2012image,xu2014deep,jain2009natural}, we use peak signal to noise ratio (PSNR) to evaluate single-channel signal separation and deconvolution quality. A higher PSNR indicates a better reconstruction quality. PSNR is defined as:
\begin{equation} \label{eq:psnr}
\text{PSNR} = 20 \ \text{log}_{10} \left ( \frac{\text{MAX}_{I}}{\sqrt{\text{MSE}}} \right ) 
\end{equation}
where $ \text{MAX}_{I} $ is the maximum value of a noise-free image. MSE represents mean squared error between two images $ I $ and $ J $ with size of $ m \times n $:
\begin{equation} \label{eq:mse}
\text{MSE}(I, K) = \frac{1}{mn} \sum_{i=0}^{m-1} \sum_{j=0}^{n-1}(I(i,j) - J(i,j))^{2}
\end{equation}

\begin{table}[t]
\centering
\begin{tabular}{*{4}{c}}
 \toprule
  & denoising & inpainting & completion \\
 \midrule
 CNN & \textbf{26.0 dB} & 15.3 dB & 12.2 dB \\
 NMF & 17.4 dB & 13.4 dB & 12.9 dB \\
 convolutive NMF & 18.3 dB & 13.4 dB & 13.0 dB \\
 S-D with 1 init. & 23.1 dB & 15.2 dB & 13.6 dB \\
 S-D with 8 init. & 25.1 dB & 18.2 dB & \textbf{15.4 dB} \\
 S-D with 32 init. & 25.1 dB & \textbf{18.9 dB} & 15.4 dB \\
 \bottomrule
\end{tabular}
\caption{PSNR of image denoising, inpainting and completion with different approaches. }
\label{table:denoise_inpaint_completion}
\end{table}

\subsection{Denoising, Inpainting and Completion}
Denoising, inpaining and completion are special case of single-channel signal separation and deconvolution problem where $ \alpha_{k} $ is an unknown constant and $ n $ is unknown noise such as Gaussian noise, non-stationary noise or corruption of an image. The first and second rows of Fig. \ref{fig:gaussian_bar_half} show the clean and noisy images. The third to the fifth rows show the denoised images with CNN, NMF and the proposed S-D approach. In the first column, testing noise and training noise have the same distribution so CNN performs well. However CNN based denoising methods do not generalize well to unseen noise such as non-stationary noise or image corruption shown in the second and third columns in Fig. \ref{fig:gaussian_bar_half}. NMF performs better than CNN under unseen noise but sometimes produces unnatural separation result, which is due to Problem 3 we stated in Section 2. S-D approach has a good performance in all of image denoising, inpainting and completion. Table \ref{table:denoise_inpaint_completion} shows PSNR of CNN, NMF, convolutive NMF and S-D approaches. S-D approach achieves a PSNR of 25.1 dB in image denoising which is comparable to CNN. NMF and convolutive NMF achieve similar PSNR of 17.4 dB and 18.3 dB, respectively. In image inpainting, S-D achieves a PSNR of 18.9 dB, outperforming NMF and CNN methods of 13.4 dB and 15.3 dB, respectively. This result shows source separation with S-D generalize well to unseen noise than NMF and CNN. In image completion, S-D approach achieves a PSNR of 15.4 dB, outperforming convolutive CNN of 12.2 dB and convolutive NMF of 12.9 dB respectively. Table \ref{table:denoise_inpaint_completion} also shows the decomposition in S-D approach with respect to the number of initializations. With 8 or 32 initializations the performance is 2 dB better than with only 1 initialization. This may result from the fact that the optimization problem in (\ref{eq:final}) is non-convex. Algorithm \ref{alg:decomposition} is a gradient based method which may lead to the solution being in a local minimum. Repeating Algorithm \ref{alg:decomposition} several times with different initializations and choosing the solution with least reconstruction error shows better performance. 

\begin{table}[t]
\centering
\begin{tabular}{cccc}
 \toprule
  & deconv. & sep. & sep. + deconv. \\
 \midrule
 NMF & 15.3 dB & 9.4 dB & 8.7 dB \\
 convolutive NMF & 18.3 dB & 14.2 dB & 10.1 dB \\
 S-D with 1 init. & 17.3 dB & 13.7 dB & 9.3 dB \\
 S-D with 8 init. & 21.9 dB & 16.8 dB & 11.5 dB \\
 S-D with 32 init. & \textbf{23.2} dB & \textbf{18.5} dB & \textbf{13.2} dB \\
 \bottomrule
\end{tabular}
\caption{PSNR of image separation and deconvolution with different approaches. }
\label{table:separation_deconvolution}
\end{table}

\begin{figure*}[t] 
  \centering
  \centerline{\includegraphics[width=0.9\textwidth]{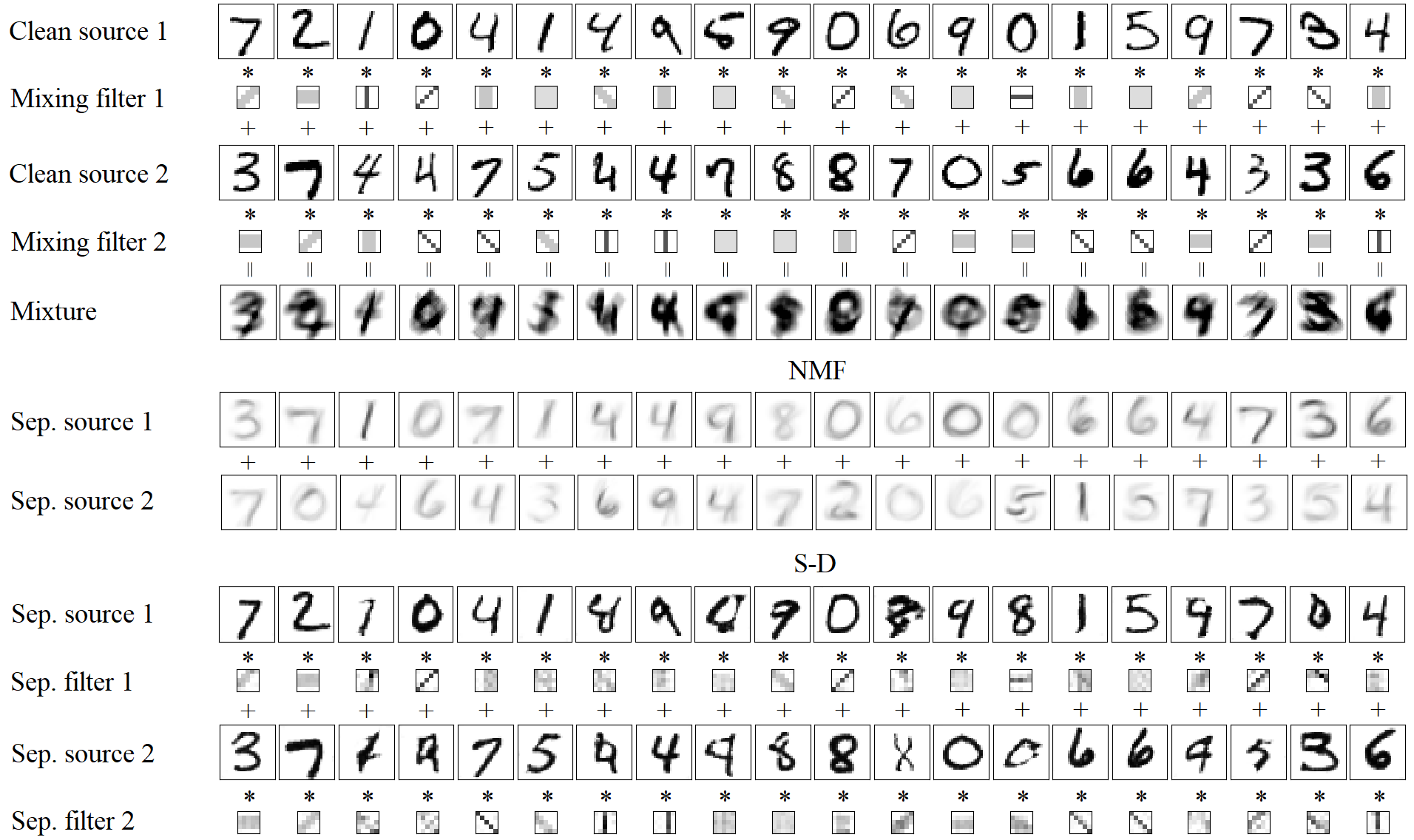}}
  \caption{Image separation and deconvolution with NMF and S-D approach. }\label{fig:mix_convolve_all}
\end{figure*}

\subsection{Separation and Deconvolution}
We evaluate single-channel signal separation and deconvolution with the mixing filters $ \alpha_{k}, k=1, ..., K $ as unknown tensors, which is a very challenging task. In this case both of the mixing tensors $ \alpha_{k} $ and individual sources $ s_{k} $ need to be estimated. Fig. \ref{fig:mix_convolve_all} shows a mixture obtained by convolving clean sources with mixing filters followed by summation. In our experiment we set $ K=2 $ and each mixing filter has a size of $ 5 \times 5 $. In actual application scenarios the size of mixing filter depends on the task. Fig. \ref{fig:mix_convolve_all} shows NMF based separation often leads to unnatural images. The S-D based approach can separate images with high quality and both the sources $ s_{k} $ and mixing filters $ \alpha_{k} $ can be estimated. Fig. \ref{fig:mix_convolve_all} shows both estimated sources and mixing filters are learned correctly compared with the ground truth sources and mixing filters. The first column of Table \ref{table:separation_deconvolution} shows the results of image deconvolution without separation where K=1 and $ \alpha $ is an unknown tensor. S-D achieves a PSNR of 23.2 dB and performs better than NMF and the convolutive NMF of 15.3 dB and 18.3 dB, respectively. The second column of Table \ref{table:separation_deconvolution} shows the results of image separation where $ \alpha_{k} $ are unknown constants and $ K=2 $. S-D achieves a PSNR of 18.5 dB and performs better than NMF and convolutive NMF of of 9.4 dB and 14.2 dB, respectively. The third column of Table \ref{table:separation_deconvolution} shows both of source separation and deconvolution where $ \alpha_{k} $ are unknown tensors and $ K=2 $. S-D achieves a PSNR of 13.2 dB and outperforms NMF and convolutive NMF of 8.7 dB and 10.1 dB, respectively. S-D with 32 initializations has higher PSNR than 8 initializations and than 1 initialization, which shows the effectiveness of repeating Algorithm \ref{alg:decomposition} several times to solve the non-convex optimization problem in (\ref{eq:final}).

\section{Conclusion}
In this paper, we propose a synthesis-decomposition (S-D) approach to solve single-channel signal separation and deconvolution problem. In synthesizing, a generative model for source signals is trained using a generative adversarial network (GAN). In decomposition, both sources and filters are optimized to minimize the reconstruction error. Instead of optimizing sources directly, we optimize over the seeds of a GAN. The proposed S-D approach achieves a PSNR of 18.9 dB and 15.4 dB in image inpainting and completion, outperforming the regression approach CNN and decomposition approach NMF. The S-D approach achieves a PSNR of 13.2 dB in image source separation with deconvolution, outperforming NMF of 8.7 dB. Repeating the decomposition in S-D several times can significantly improve PSNR. In future, we will explore the S-D approach to more source separation and deconvolution problems. 

\section*{Acknowledgements}
This research was supported by EPSRC grant EP/N014111/1 ``Making Sense of Sounds'' and a Research Scholarship from the China Scholarship Council (CSC) No. 201406150082.

\bibliographystyle{named}
\bibliography{ijcai19}

\end{document}